# On the correlation between bibliometric indicators and peer review: Reply to Opthof and Leydesdorff

Ludo Waltman, Nees Jan van Eck, Thed N. van Leeuwen,
Martijn S. Visser, and Anthony F.J. van Raan

Centre for Science and Technology Studies, Leiden University, The Netherlands
{waltmanlr, ecknjpvan, leeuwen, visser, vanraan}@cwts.leidenuniv.nl

Opthof and Leydesdorff [arXiv:1102.2569] reanalyze data reported by Van Raan [arXiv:physics/0511206] and conclude that there is no significant correlation between on the one hand average citation scores measured using the CPP/FCSm indicator and on the other hand the quality judgment of peers. We point out that Opthof and Leydesdorff draw their conclusions based on a very limited amount of data. We also criticize the statistical methodology used by Opthof and Leydesdorff. Using a larger amount of data and a more appropriate statistical methodology, we do find a significant correlation between the CPP/FCSm indicator and peer judgment.

## 1. Introduction

In this note, we reply to a recent contribution by Opthof and Leydesdorff entitled "Citation analysis cannot legitimate the strategic selection of excellence" (Opthof & Leydesdorff, 2011; see also Opthof & Leydesdorff, in press; henceforth O&L). Although O&L present their contribution as a comment to one of our recent papers (Waltman, Van Eck, Van Leeuwen, Visser, & Van Raan, 2011), their contribution in fact focuses almost completely on an earlier paper written by one of us (Van Raan, 2006).

Van Raan (2006) considers 147 Dutch research groups in chemistry and studies how two bibliometric indicators, namely the *h*-index (Hirsch, 2005) and the CPP/FCSm indicator, correlate with the quality judgment of a peer review committee. Based on the data reported by Van Raan (Tables 1 and 2), O&L reanalyze the correlation of the two bibliometric indicators with peer judgment. O&L conclude that there is no significant correlation between the CPP/FCSm indicator and peer judgment. They also conclude that the CPP/FCSm indicator fails to distinguish between 'good' and 'excellent' research.

Below, we comment on the statistical analysis of O&L. We also make a more general remark on the comparison of citation analysis and peer review.

## 2. Data

The analysis of Van Raan (2006) is based on an assessment study of Dutch chemistry and chemical engineering research groups conducted by the Association of Universities in the Netherlands (for a full description of the study, see VSNU, 2002). For each research group, our institute, the Centre for Science and Technology Studies of Leiden University, calculated a number of bibliometric indicators (see our report included at the end of VSNU, 2002). One of the indicators is the CPP/FCSm indicator. This indicator measures a research group's average number of citations per publication, where citations are normalized for differences among fields. The assignment of publications to researchers was verified by the researchers themselves.

In the original study, the CPP/FCSm indicator was calculated based on publications from the period 1991–2000. However, the analysis of Van Raan only uses publications from the period 1991–1998. Our analysis presented below uses the same data as the analysis of Van Raan.

The peer review committee, which consisted of eleven members, assessed the research groups on four dimensions: Scientific quality, scientific productivity, scientific relevance, and long term viability. For each research group, the committee provided both a written appraisal and numerical scores. A separate numerical score was given for each of the four above-mentioned dimensions. Numerical scores were given on a five-point scale: 1 ('poor'), 2 ('unsatisfactory'), 3 ('satisfactory'), 4 ('good'), and 5 ('excellent'). The bibliometric indicators calculated by our institute were provided to the committee members before the start of the peer review procedure. This means that the bibliometric indicators may have influenced the judgments of the peer review committee.

The analysis of Van Raan (2006) focuses on the numerical scores given by the peer review committee on the dimension of scientific quality. For some research groups, a quality score is not available. These research groups are excluded from the analysis. There are 147 research groups for which a quality score is available. None of these groups has a score of 1 or 2. Hence, all groups have a score of 3 (30 groups), 4 (78 groups), or 5 (39 groups). The average number of publications used in the calculation of the CPP/FCSm score of a research group is 140.

To allow others to verify our analysis presented below, the CPP/FCSm scores and the quality scores of the 147 research groups have been made available online. The scores can be downloaded from www.cwts.nl/research/bibliometrics_vs_peer_review/data.txt.

## 3. Analysis

Based on the data reported by Van Raan (2006) in Tables 1 and 2 of his paper, O&L draw the following conclusions:

1. There is no significant correlation between the CPP/FCSm indicator and the quality judgment of the peer review committee.
2. The CPP/FCSm indicator performs poorly in distinguishing between 'good' and 'excellent' research.

In our view, O&L base their conclusions on a flawed statistical analysis. We have two important objections against the statistical analysis of O&L. First, the statistical analysis is based on a very limited amount of data. O&L did not have access to the full data set used by Van Raan (2006), and they therefore based their analysis on the data reported by Van Raan in his paper (in Tables 1 and 2). As a consequence, the first conclusion of O&L mentioned above is based on only 12 observations. The second conclusion is based on 147 observations, but in this case CPP/FCSm scores have been reduced to three CPP/FCSm ranges (i.e., CPP/FCSm below 1, CPP/FCSm between 1 and 2, and CPP/FCSm above 2). Clearly, reducing CPP/FCSm scores to three CPP/FCSm ranges causes a large loss of information.

Our second objection against the statistical analysis of O&L is more fundamental. Even if the analysis of O&L had been based on a much larger amount of data, their statistical methodology would not have been appropriate to determine the degree to which the CPP/FCSm indicator correlates with the quality judgment of the peer review committee. The methodology of O&L, which relies on statistical hypothesis testing, is focused entirely on determining whether a relation between the CPP/FCSm indicator and peer judgment can be established. However, with a sufficiently large amount of data, it will almost always be possible to establish such a relation. What is much more important, in our view, is to focus on the strength of the relation between the CPP/FCSm indicator and peer judgment (rather than on the artificial dichotomy between the presence and the absence of a relation).[1]

Using a more appropriate statistical methodology, we now investigate the validity of the conclusions drawn by O&L. We use the full data set of Van Raan (2006).

Table 1 reports the median, the mean, and the standard deviation of the CPP/FCSm scores of the 147 research groups. The results are reported both for all research groups together and separately for the research groups with a quality score of 3 ('satisfactory'), 4 ('good'), or 5 ('excellent'). The table also reports a 95% confidence interval for the mean of the CPP/FCSm scores.[2] Figures 1 and 2 provide box plots and a histogram that show the distribution of the CPP/FCSm scores over the research groups.

Table 1. Descriptive statistics for the CPP/FCSm scores of the 147 research groups.

| Quality score | No. of research groups | Median CPP/FCSm | Mean CPP/FCSm | St. dev. CPP/FCSm | 95% conf. int. mean CPP/FCSm |
|---|---|---|---|---|---|
| 3 | 30 | 1.04 | 1.02 | 0.45 | 0.87–1.19 |
| 4 | 78 | 1.45 | 1.55 | 0.64 | 1.41–1.69 |
| 5 | 39 | 1.81 | 1.99 | 0.84 | 1.74–2.26 |
| All | 147 | 1.39 | 1.56 | 0.74 | 1.44–1.68 |

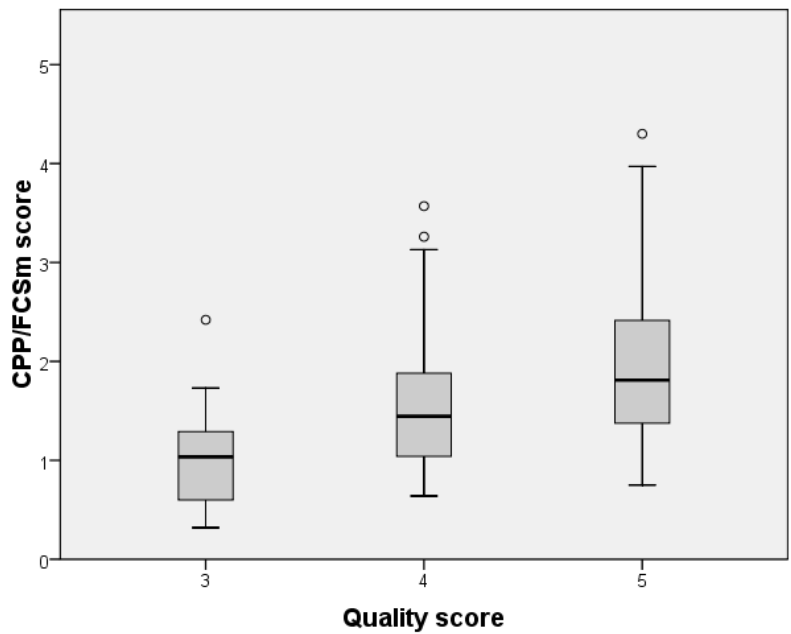

Figure 1. Box plots showing the distribution of the CPP/FCSm scores over the research groups. A separate box plot is provided for each quality score.

---

[1] Statistical hypothesis testing has many limitations and problems, and its extensive use in the social sciences is often criticized. For an introduction into the literature on this issue, see for example Kline (2004).

[2] All confidence intervals that we report were calculated using a bootstrapping approach (e.g., Efron & Tibshirani, 1993).

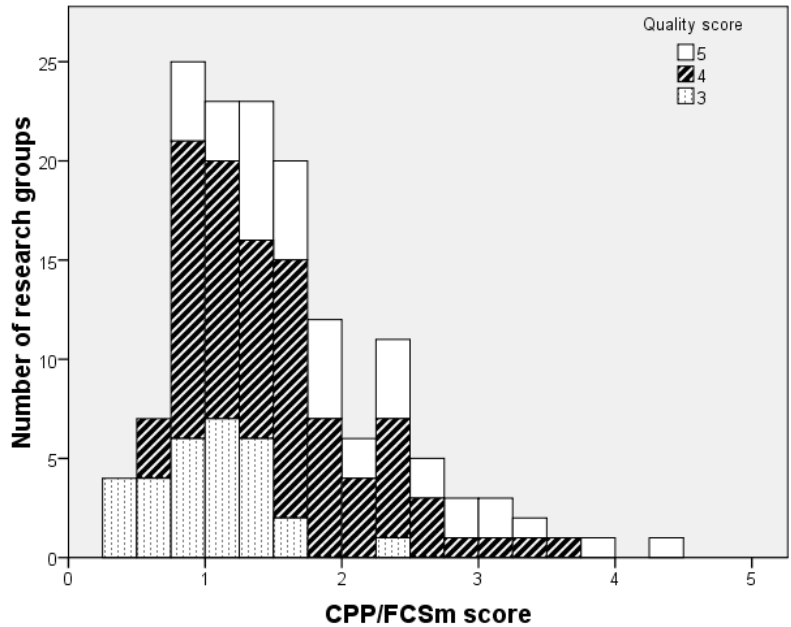

Figure 2. Histogram showing the distribution of the CPP/FCSm scores over the research groups. Shading is used to indicate the quality scores of the research groups.

In Table 1 and Figure 1, we observe that on average research groups with a quality score of 5 have a substantially higher CPP/FCSm score than research groups with a quality score of 4, while the latter research groups in turn have a substantially higher CPP/FCSm score than research groups with a quality score of 3. The difference in mean CPP/FCSm score between research groups with a quality score of 5 and research groups with a quality score of 4 is 0.44 (95% conf. int.: 0.15–0.74). For research groups with a quality score of 4 and research groups with a quality score of 3, the difference is 0.53 (95% conf. int.: 0.31–0.73).[3] Clearly, the observed differences are significant not only from a statistical point of view but also from a substantive point of view. We therefore conclude that the CPP/FCSm indicator is significantly correlated with the quality judgment of the peer review committee. This contradicts the first conclusion of O&L mentioned above.

The Spearman rank correlation between CPP/FCSm scores and quality scores equals 0.45 (95% conf. int.: 0.31–0.57), which suggests a moderately strong correlation.[4] This is in line with Figures 1 and 2. The figures show that research groups with a quality score of 3 and research groups with a quality score of 4 are fairly well separated from each other in terms of their CPP/FCSm scores. However, consistent with results reported by Moed (2005, Chapter 19), the separation between research groups with a quality score of 4 and research groups with a quality score of 5 is not so good. O&L conclude that the CPP/FCSm indicator performs poorly in distinguishing between these two quality scores. In our view, this conclusion is too strong, given the fact that research groups with a quality score of 5 on average have an almost 30% higher CPP/FCSm score than research groups with a quality score of 4 (1.99 vs. 1.55; see Table 1).

---

[3] For comparison, suppose the 147 research groups would be sorted in increasing order of their CPP/FCSm score, and suppose the first 30 groups would be given a quality score of 3, the next 78 groups would be given a quality score of 4, and the final 39 groups would be given a quality score of 5. The mean CPP/FCSm scores of the groups with a quality score of 3, 4, and 5 would then be 0.75, 1.37, and 2.55, respectively. Hence, for groups with a quality score of 5 and groups with a quality score of 4, the difference would be 1.18 (rather than 0.44). For groups with a quality score of 4 and groups with a quality score of 3, the difference would be 0.62 (rather than 0.53).

[4] The correlation of 0.45 is somewhat lower than the correlations reported by Moed (2005, p. 241) for a number of similar data sets. It should be noted that because of the many ties in the quality scores it is impossible to obtain a Spearman rank correlation of one. A more appropriate correlation measure would be the variant of the Kendall rank correlation discussed by Adler (1957). Using this measure, it is always possible to obtain a correlation of one. We obtain a correlation of 0.46 (95% conf. int.: 0.32–0.59) using this measure.

## 4. Citation analysis versus peer review

Finally, we want to make a more general remark on the comparison of citation analysis and peer review. Based on their analysis, O&L conclude that bibliometric indicators have difficulties in distinguishing between good and excellent research. However, this conclusion rests on an important implicit assumption, namely the assumption that the peer review committee has been able to distinguish between good and excellent research with a high degree of accuracy. This is a strong assumption. There is an extensive literature which indicates that peer review, just like citation analysis, has significant limitations (for an overview, see Bornmann, 2011). For instance, many studies report a relatively low reliability of peer review, and peer review is also often suggested to suffer from various types of biases. Given the limitations of both citation analysis and peer review, discrepancies between the two can always be interpreted in two directions. Based on our analysis presented above, it may be that bibliometric indicators indeed have difficulties in distinguishing between good and excellent research. However, it may also be that instead of the indicators the peers have difficulties in making this distinction (as suggested by Moed, 2005, Chapter 19 and Rinia, Van Leeuwen, Van Vuren, & Van Raan, 1998). O&L ignore this second possibility and seem to assume that discrepancies between citation analysis and peer review can only be explained in terms of shortcomings of the bibliometric indicators. In our view, this is a too simplistic perspective on the intricate relation between citation analysis and peer review.